\documentclass[a4paper, oneside, twocolumn, notitlepage, 10pt]{extarticle_ecoc}
\usepackage{ecoc}
\usepackage{subcaption}
\usepackage{graphicx}
\usepackage{setspace}
\usepackage{lipsum}  
\usepackage{layouts}
\usepackage{xcolor}
\usepackage{amsmath,mathtools}
\usepackage{tikz}
\usepackage{pgfplots}
\usetikzlibrary{pgfplots.groupplots}
\pgfplotsset{compat=newest}

\begin{document}
\selectlanguage{english}    


\title{End-to-End Deep Learning of Long-Haul Coherent Optical Fiber Communications via Regular Perturbation Model}

%

\author{
    Vladislav Neskorniuk\textsuperscript{(1,2)}, Andrea Carnio\textsuperscript{(3)},
    Vinod Bajaj\textsuperscript{(1,4)}, Domenico Marsella\textsuperscript{(3)},\\Sergei K. Turitsyn\textsuperscript{(2)}, Jaroslaw E. Prilepsky\textsuperscript{(2)}, Vahid Aref\textsuperscript{(1)}
}

\maketitle                  


\begin{strip}
 \begin{author_descr}

   \textsuperscript{(1)}Nokia, 70435 Stuttgart, Germany,
   \textcolor{blue}{\uline{v.neskorniuk@aston.ac.uk}}, \textcolor{blue}{\uline{vahid.aref@nokia.com}}

   \textsuperscript{(2)}Aston Institute of Photonic Technologies, Aston University, B4 7ET Birmingham, UK

   \textsuperscript{(3)}Nokia, Vimercate 20871, Italy
   
   \textsuperscript{(4)}Delft Center for Systems and Control, Delft University of Technology, 2628 CD Delft, The Netherlands

 \end{author_descr}
\end{strip}

\setstretch{1.1}
\renewcommand\footnotemark{}
\renewcommand\footnoterule{}


\begin{strip}
  \begin{ecoc_abstract}
    We present a novel end-to-end autoencoder-based learning for coherent optical communications using a ``parallelizable'' perturbative channel model. We jointly optimized constellation shaping and nonlinear pre-emphasis achieving mutual information gain of 0.18 bits/sym./pol. simulating 64 GBd dual-polarization single-channel transmission over 30x80 km G.652 SMF link with EDFAs.
  \end{ecoc_abstract}
\end{strip}

\section{Introduction}

\begin{figure*}[t!]
    \centering
    \includegraphics[width=\textwidth]{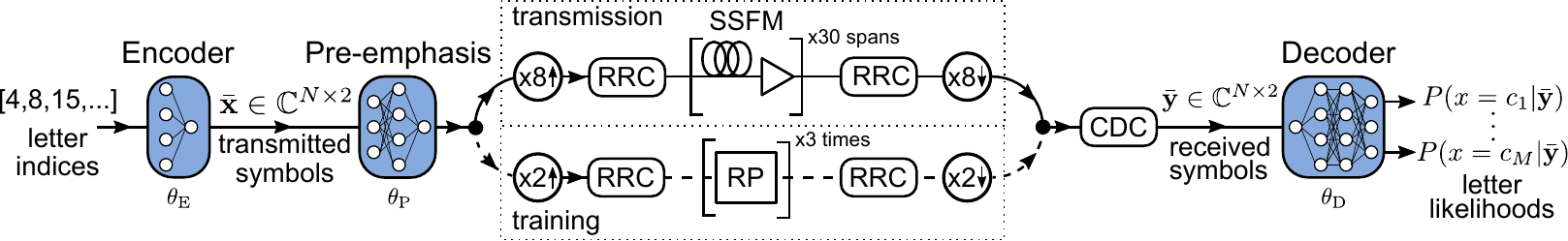}
    \caption{Principal scheme of the implemented autoencoder, trained over auxiliary RP model and assessed over SSFM simulation.}
    \label{fig:ae-scheme}
\end{figure*}
\begin{figure}[t]
    \centering
    \includegraphics[width=0.48\textwidth]{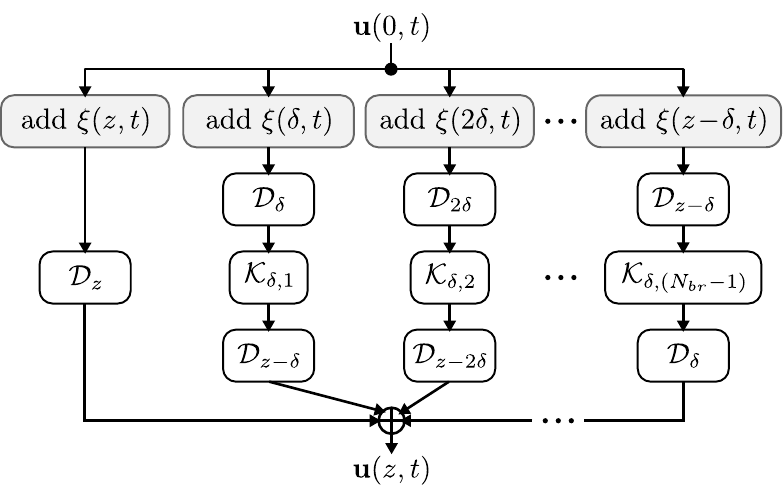}
    \caption{Principal scheme of the first-order regular-perturbation (RP) algorithm. 
    }
    \label{fig:rp-scheme}
\end{figure}
%

In modern communication systems, transceivers typically contain a chain of digital signal processing (DSP) blocks individually designed based on analytical models. 
The end-to-end (E2E) neural network (NN)-based \textit{autoencoders} have
become of particular interest to improve the overall system performance, particularly for 
the scenarios where accurate models are either unknown or computationally prohibitive to use.
In this approach, the transmitter (TX), the 
channel, and the receiver (RX) are implemented as a single deep NN, and then, TX and RX are jointly trained to reproduce the TX inputs from the RX outputs.
 
The autoencoder-based communication system design has been first proposed~\cite{IntroDeepLearningPhyLayer} and, subsequently, realized for various communication systems~\cite{dorner2017deep,cammerer2020trainable,stark2019joint,karanov2018JLTe2e,karanov2020concept,gumucs2020GMIGeoShape,RajoECOC2018,Simone,li2018achievable,song2021end}. 
In optical fiber communications, the E2E learning has been applied for both intensity modulation and direct detection (IM/DD) systems~\cite{karanov2018JLTe2e,karanov2019classicalDSP,karanov2020concept} and coherent systems~\cite{RajoECOC2018,Simone,li2018achievable,uhlemann2020deep};  
For the latter, E2E learning is much more involved. 
The nonlinear dispersive channel is typically modeled by the Manakov equation and simulated by a \textit{serial} cascade of alternating linear and nonlinear operators, known as the split-step Fourier method (SSFM)~\cite{agrawal2000nonlinear}.
The corresponding neural network consists of many layers making the training process via ``back-propagation'' very slow and challenging. 
It requires the storage of all intermediate states, thus, making the process memory hungry. In addition, the back-propagation through many layers is prone to uncontrolled growth or vanishing of gradients~\cite{uhlemann2020deep}. 
To bypass these problems, 
the one way is to approximate the channel with simplified models. For instance, the E2E learning is done using dispersion-free channel model~\cite{li2018achievable} or Gaussian noise model~\cite{RajoECOC2018} which considers nonlinear distortion as an additive noise; 
However, these two models do not account for channel memory. 

In this paper, we propose E2E learning via the first-order regular perturbation (RP) model. As we will show, RP model offers not only a quite accurate approximation of the Manakov equation in the power range of interest but, also, it can be implemented in \textit{parallel} branches, an architecture suitable for neural networks optimization. As a case study, we consider single-channel dual-polarization (DP) 64 Gbaud transmission over 30 spans of 80 km standard single mode fiber (SSMF) with lumped optical amplifiers (OAs). We assume the linear coherent reception without any nonlinear equalization. 
For a range of launch powers, we learn optimized 64-point geometrically shaped (GS-64) constellations and nonlinear pre-emphasis filters maximizing the E2E mutual information (MI). 
The training is done via RP model but the performance is evaluated over SSFM (as a precise channel model). 
We show that in comparison to the standard 64-QAM, the learned GS-64 constellation and waveforms increase the optimal launch power by about 0.5~dB and improve the MI from 4.95 bits/sym./pol. to 5.13 bits/sym./pol. 

\section{RP as Auxiliary  Channel Model}
Consider the Manakov equation describing a DP optical signal $\mathbf{E}(z,t)=\mathbf{u}(z,t)\sqrt{f(z)}$ over a fiber-optic link with lumped amplification~\cite{agrawal2000nonlinear, mecozzi2012nonlinear}. $f(z)=\exp\left( -\alpha z + \alpha L_\text{sp} \lfloor z/L_\text{sp}\rfloor \right)$ models the optical losses~$\alpha$ and amplification, $L_\text{sp}$ is the fiber span length
\begin{equation*}
    \frac{\partial \mathbf{u}}{\partial z} = - i\frac{\beta_2}{2}\frac{\partial^2\mathbf{u}}{\partial t^2} + i\frac{8}{9}\gamma f(z)\| \mathbf{u}\|^2 \mathbf{u} +\eta(z,t),
\end{equation*}
 where $\beta_2$ and $\gamma$ are the dispersion and Kerr nonlinearity coefficients; $\eta(z,t)$ denotes the amplified spontaneous emission noise (ASE) of OAs.

The first-order regular perturbation (RP) is an elaborate method to approximate $\mathbf{u}(z,t)$ in a weakly nonlinear regime as~\cite{vannucci2002rp, mecozzi2012nonlinear, garciagomez2021mismatched}
\begin{align*}
    \mathbf{u}(z,t) &= \mathbf{u}_{\rm L}(z,t) + \mathbf{u}_{\rm NL}(z,t) + \mathcal{O}(\gamma^2),\\
    \mathbf{u}_{\rm L}(z,t) &= \mathcal{D}_z\left[\mathbf{u}(0,t) + \eta(z,t)\right],\\
    \mathbf{u}_{\rm NL}(z,t) &\approx \sum_{m=1}^{N_{\rm br}-1}\mathcal{D}_{z-m\delta}\left[\mathcal{K}_{\delta,m}[\mathbf{u}_{\rm L}(m\delta,t)]\right],\\
    \mathcal{K}_{\delta,m}[\mathbf{u}(t)] &= i\frac{8}{9}\gamma \frac{1-e^{-\alpha \delta}}{\alpha} f(m\delta)  \|\mathbf{u}(t) \|^2 \mathbf{u}(t),
\end{align*}
where $\mathcal{D}_z[\cdot] = \mathcal{F}^{-1} \left[\exp(i\beta_2z\omega^2/2) \mathcal{F}[\cdot]\right]$ is chromatic dispersion operator, $\|\cdot\|$ is 2-norm, and $\mathcal{F}$ denotes Fourier transform. Fig.~\ref{fig:rp-scheme} shows the block diagram of the above equations. The leftmost branch gives $\mathbf{u}_{\rm L}(z,t)$, while the other branches sum to $\mathbf{u}_{\rm NL}(z,t)$. The number of branches is $N_{\rm br}=z/\delta$.
The smaller $\delta$ is, the more accurately $\mathbf{u}_{\rm NL}(z,t)$ can be approximated.
Each branch also includes an additive circularly-symmetric weight Gaussian noise $\xi(z',t)$ with power spectral density $\lfloor\frac{z'}{L_{\rm sp}}\rfloor\sigma_{\rm ASE}^2$.
A link can be modeled by just a single stage ($z$ is the link length) or few subsequent stages of the RP model.
It is evident that a stage of the RP model is easily parallelizable, i.e. all branches can be computed independently and in parallel. This allows speeding up its calculation and so the overall E2E learning by using graphics processing units (GPUs). Moreover, the danger of exploding or vanishing gradients is reduced, compared to sequential models like SSFM.
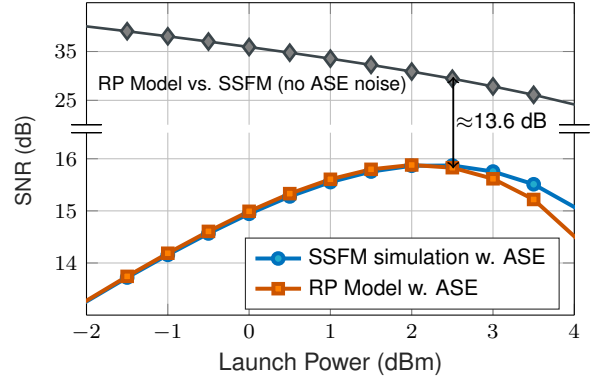
\begin{figure}[t]
  \centering
    \pgfplotsset{
    every non boxed x axis/.style={font=\footnotesize}
}
\pgfplotsset{every axis label/.append style={font=\footnotesize}}
\pgfplotsset{every tick label/.append style={font=\footnotesize}}
\pgfplotsset{every legend label/.append style={font=\footnotesize}}

\definecolor{mycolor1}{rgb}{0.00000,0.44700,0.74100}%
\definecolor{arsenic}{rgb}{0.23, 0.27, 0.29}
\definecolor{mycolor2}{rgb}{0.00000,0.44706,0.74118}%
\definecolor{mycolor3}{rgb}{0.49412,0.18431,0.55686}%
\definecolor{mycolor4}{rgb}{0.50196,0.50196,0.50196}%
\definecolor{ballblue}{rgb}{0.13, 0.67, 0.8}
\definecolor{byzantine}{rgb}{0.74, 0.2, 0.64}
\definecolor{darkbyzantium}{rgb}{0.36, 0.22, 0.33}
\definecolor{darkmagenta}{rgb}{0.55, 0.0, 0.55}
\definecolor{burntorange}{rgb}{0.8, 0.33, 0.0}
\begin{tikzpicture}
\begin{groupplot}[
    group style={
        group name=my fancy plots,
        group size=1 by 2,
        xticklabels at=edge bottom,
        vertical sep=3pt
    },
    at={(0,0)},
    xmin=-2,
    xmax=4,
    width=0.5\textwidth,
    xmajorgrids,
    ymajorgrids, 
    xlabel style={font=\footnotesize \color{white!15!black},at={(0.5,-2ex)}},
    ylabel style={font=\footnotesize \color{white!15!black},at={(-3ex,0.5)}},
    legend style={font=\footnotesize, legend cell align=left, align=left,legend pos=south east, draw=white!15!black}
]
\nextgroupplot[ymin=20,ymax=45,
               ytick={25,35},
               axis x line=top,
               height=0.2\textwidth]
\addplot [color=arsenic, line width=1.0pt, mark size=3pt, mark=diamond*, mark options={solid, fill=mycolor4}]
  table[row sep=crcr]{%
-5	45.4564841817944\\
-4.5	44.6724329265286\\
-4	43.8444995471789\\
-3.5	42.9764443121199\\
-3	42.0714900353434\\
-2.5	41.1320711537833\\
-1.5	39.1544472551584\\
-1	38.1154716944791\\
-0.5	37.0401520878385\\
0	35.9242306534243\\
0.5	34.7615000018132\\
1	33.5435151760114\\
1.5	32.2592709168173\\
2	30.894874632254\\
2.5	29.4332775242649\\
3	27.8541712884197\\
3.5	26.1342154327389\\
4.5	22.1688903576412\\
5	19.8738330801166\\
};
\coordinate(dtl)at(rel axis cs:0,0);
\coordinate(dtr)at(rel axis cs:1,0);
\node [scale=0.8] at (rel axis cs: 0.34,0.33) 
{\small RP Model vs. SSFM (no ASE noise)};

\nextgroupplot[ymin=13,ymax=16.5,
               ytick={14,15,16},
               axis x line=bottom,
               xlabel style={font=\color{white!15!black}},
               xlabel={\small Launch Power (dBm)},
               ylabel=\qquad \qquad \qquad {SNR (dB)},
               height=0.25\textwidth]

\addplot [color=mycolor1, line width=1.5pt, mark size=2.0pt, mark=*, mark options={solid, fill=ballblue}]
  table[row sep=crcr]{%
-5	10.3715448640961\\
-4.5	10.8649450478265\\
-4	11.3558173153964\\
-3.5	11.8431684040553\\
-3	12.3256166603518\\
-2.5	12.80124616042\\
-1.5	13.7204923763575\\
-1	14.1555681608474\\
-0.5	14.5660629613577\\
0	14.9433535253153\\
0.5	15.2764450455307\\
1	15.5518504693098\\
1.5	15.7538852710365\\
2	15.8656099064821\\
2.5	15.8705504819466\\
3	15.7550439971409\\
3.5	15.5106509861416\\
4.5	14.6360144827983\\
5	14.0225305759928\\
};
\addlegendentry{\footnotesize SSFM simulation w. ASE}

\addplot [color=burntorange, line width=1.5pt, mark size=1.8pt, mark=square*, mark options={solid, fill=orange, draw = burntorange}]
  table[row sep=crcr]{%
-5	10.3603305493124\\
-4.5	10.8563175430355\\
-4	11.3503699458285\\
-3.5	11.8416058784831\\
-3	12.3287612903782\\
-2.5	12.810031090812\\
-1.5	13.7435887999748\\
-1	14.1871645306449\\
-0.5	14.6065235127234\\
0	14.9920346129729\\
0.5	15.3308299018559\\
1	15.6062345666143\\
1.5	15.7975348837417\\
2	15.8804329002989\\
2.5	15.828502724459\\
3	15.6156362961871\\
3.5	15.218811952837\\
4.5	13.805007007457\\
5	12.76179920509\\
};
\addlegendentry{\footnotesize RP Model w. ASE}
\coordinate(dbl)at(rel axis cs:0,1);
\coordinate(dbr)at(rel axis cs:1,1);

\end{groupplot}

\draw(dtl)--+(-0.2,0)--+(0.2,0)--+(0,0);
\draw(dbl)--+(-0.2,0)--+(0.2,0)--+(0,0);
\draw(dtr)--+(-0.2,0)--+(0.2,0)--+(0,0);
\draw(dbr)--+(-0.2,0)--+(0.2,0)--+(0,0);
\draw[arrows=<->,line width=0.7pt](1.9in,-0.228in)--(1.9in,0.25in);
\node[] at (2.15in,0.02in) {\footnotesize{$\approx$13.6 dB}};

\end{tikzpicture}
    \caption{ Comparison of 3-stages RP model with the SSMF simulation. The approximation error of the RP model is much smaller than the total distortion. 
    }
    \label{fig:rp_vs_ssfm}
\end{figure}

Let us now discuss the accuracy of RP. As a testcase, we numerically consider single-channel DP 64-QAM transmission at 64 Gbaud over 30 spans of 80 km SSMF. A root-raised-cosine with roll-off factor of 0.1 is used for pulse shaping. Manakov equation was used as a reference channel model. We define the SSMF parameters as: $\alpha = 0.21$ dB/km, $\beta_2 = -21.4$ $\text{ps}^2$/km, and $\gamma = 1.14$~$\text{(W*km)}^{-1}$. Every span was followed by an ideal lumped OA with noise figure $\text{NF}=4$~dB.

To have a more accurate model, we used 3 subsequent stages of RP, 
each covering 10 spans with $N_{\rm br}=100$.
We compare this auxiliary channel model to the precise SSFM in Fig.~\ref{fig:rp_vs_ssfm}. 
We plot the signal-to-noise ratio (SNR) of the received signals after chromatic dispersion compensation (CDC), as depicted in Fig.~\ref{fig:ae-scheme}. We see that the received SNR is quite similar for both SSFM and 3-stages RP model in weak nonlinear regime (up to 2.5~dBm). To illustrate the approximation error of RP, we compare the outputs of our RP model $\mathbf{\bar{y}}_{\rm RP}$ with the outputs of SSFM $\mathbf{\bar{y}}_{\rm SSFM}$ with no additional ASE noise. We characterize the approximation error in terms of signal-to-distortion ratio (SDR), defined as $-20\log_{10}\left(\|\mathbf{\bar{y}}_{\rm RP}-\mathbf{\bar{y}}_{\rm SSFM}\|/\|\mathbf{\bar{y}}_{\rm SSFM}\|\right)$. We see in Fig.~\ref{fig:rp_vs_ssfm} that up to 2.5 dBm, the SDR is at least 13 dB larger than the received SNR, implying that the approximation error of the RP model is much smaller than the total distortion in the link.

\vspace{-1mm}
\section{E2E Learning Procedure and the Results}

Fig.~\ref{fig:ae-scheme} illustrates the proposed E2E autoencoder including three separate neural networks:
\\
\underline{\textit{Encoder NN}}: It is a single linear layer with trainable weights $\theta_{\rm E}$, which maps a one-hot vector of size 64, representing the transmitted message, to the corresponding constellation point $c \in \mathbb{C}$. Constellation power is fixed $\mathbb{E}\{\|c\|^2\} = 1$.
\\
\underline{\textit{Nonlinear pre-emphasis NN}}: It is implemented based on the known cubic correction terms~\cite{tao2011multiplier, malekiha2016efficient} $\Delta x_{h/v,0} = \sum_{m,n} C_{m,n}\,x_{h/v,m}\cdot (x_{h/v,n}x^*_{h/v,m+n} + x_{v/h,n}x^*_{v/h,m+n})$, where ${x}_{h/v,m}$ is the $m$-th adjacent symbol in H-/V-polarization of target symbol ${x}_{h/v,0}$. The trainable weights $\theta_{\rm P}=\{C_{m,n}\}$ with $|m|\leq 10,|n|\leq 10$ are initialized according to~\cite{ghazisaeidi2014calculation}.
\\
\underline{\textit{Decoder NN}}: It is a dense NN with trainable weights $\theta_{\rm D}$ composed of size-1 complex-valued input layer followed by
2 hidden layers, 32 ReLU neurons each, and size-64 softmax as output layer activation. It maps a received symbol $y \in \mathbb{C}$ to 64 posterior probabilities $P(c_k|y)$ of each constellation point $c_k$.

Note that the TX NN was divided into two parts to reduce training complexity and improve interpretability~\cite{uhlemann2020deep,song2021end}. The same encoder and decoder NNs were applied to both polarizations.

The autoencoder is trained on the RP model to maximize the E2E mutual information (MI), i.e.,
\begin{equation*}
    I_{\rm RP}^*= 6+\max_{\theta_{\rm E},\theta_{\rm P},\theta_{\rm D}} \mathbb{E}_{X,Y}\{\log_2 P(x|y)\}\vspace{-1mm}
\end{equation*}
where the maximization objective is the negative categorical cross entropy.
Using a large random training sequence, the Adam optimizer~\cite{kingma2014adam} is used to maximize the objective function and to obtain the optimal $\theta^*_{\rm E}$, $\theta^*_{\rm P}$, $\theta^*_{\rm D}$.
Next, these learnt parameters are used to assess the performance over SSFM simulation.
To improve matching of the NN decoder to the actual channel, $\theta^*_{\rm D}$ is fine-tuned on the SSFM propagation data, maximizing the E2E MI $I_{\rm SSFM}^*$. Finally, the MI is assessed by transmission simulation of 10 newly generated random sequences of $2^{16}$ symbols over SSFM.

\begin{figure}[t]
  \centering
%
%
\pgfplotsset{
    every non boxed x axis/.style={font=\footnotesize}
}
\pgfplotsset{every tick label/.append style={font=\footnotesize}}

\definecolor{mycolor1}{rgb}{0.00000,0.44700,0.74100}%
\definecolor{arsenic}{rgb}{0.23, 0.27, 0.29}
\definecolor{mycolor2}{rgb}{0.00000,0.44706,0.74118}%
\definecolor{mycolor3}{rgb}{0.49412,0.18431,0.55686}%
\definecolor{mycolor4}{rgb}{0.50196,0.50196,0.50196}%
\definecolor{ballblue}{rgb}{0.13, 0.67, 0.8}
\definecolor{byzantine}{rgb}{0.74, 0.2, 0.64}
\definecolor{darkbyzantium}{rgb}{0.36, 0.22, 0.33}
\definecolor{darkmagenta}{rgb}{0.55, 0.0, 0.55}
\definecolor{forestgreen}{rgb}{0.0, 0.27, 0.13}
\definecolor{emerald}{rgb}{0.31, 0.78, 0.47}

\begin{tikzpicture}
\begin{axis}[%
width=0.4\textwidth,
height=0.3\textwidth,
at={(0.758in,0.481in)},
scale only axis,
xmin=0.95,
xmax=3.05,
xlabel={Launch power [dBm]},
xtick distance=0.25,
ymin=4.85,
ymax=5.15,
ylabel={MI [bits/sym./pol.]},
xlabel style={font=\footnotesize \color{white!15!black},at={(0.5,-2ex)}},
ylabel style={font=\footnotesize \color{white!15!black},at={(-4ex,0.5)}},
ytick distance=0.05,
axis background/.style={fill=white},
xmajorgrids,
ymajorgrids,
legend columns=2,
legend style={font= \footnotesize, at={(-0.09,1.05)}, anchor=south west, legend cell align=left, align=left, draw=white!15!black}
]

\addplot [color=arsenic, dashed, line width = 1.0pt, mark size= 2pt, mark=square*, mark options={solid, fill = mycolor4, draw= arsenic}]
 plot [error bars/.cd, y dir = both, y explicit,     error bar style={line width=0.8pt,solid},
      error mark options={line width=0.8pt,mark size=3pt,rotate=90}]
 table[row sep=crcr, y error plus index=2, y error minus index=3]{%
1       4.877409            0.006283            0.006283
1.25	4.90728082531652	0.00709177380329537	0.00709177380329537\\
1.5	4.93068496741442	0.00792583006080243	0.00792583006080243\\
1.75	4.94672032871379	0.00906604819591508	0.00906604819591508\\
2	4.95465029863568	0.0102301064034585	0.0102301064034585\\
2.25	4.95391777534442	0.0114886399990239	0.0114886399990239\\
2.5	4.94322645755827	0.0130386831744807	0.0130386831744807\\
2.75	4.9219395676893	0.0147016439156075	0.0147016439156075\\
3	4.8891405448698	0.0164648055359016	0.0164648055359016\\
};
\addlegendentry{64QAM, KDE}

\addplot [color=arsenic, line width = 1.0pt, mark size = 2pt, mark=square*, mark options={solid, fill = mycolor4, draw = arsenic}]
 plot [error bars/.cd, y dir = both, y explicit,error bar style={line width=0.8pt,solid},
      error mark options={line width=0.8pt,mark size=3pt,rotate=90}]
 table[row sep=crcr, y error plus index=2, y error minus index=3]{%
1       4.83483032748066    0.006798            0.006798\\
1.25	4.86581624361917	0.00766049113829521	0.00766049113829521\\
1.5	4.89092564582529	0.00848049003745889	0.00848049003745889\\
1.75	4.90993872156966	0.00916836269855365	0.00916836269855365\\
2	4.92217073724284	0.0100210733196929	0.0100210733196929\\
2.25	4.92699404076043	0.0110460523004056	0.0110460523004056\\
2.5	4.92362440419207	0.0122163497502164	0.0122163497502164\\
2.75	4.91229110968826	0.0131403055036957	0.0131403055036957\\
3	4.89200011697216	0.0143853673690917	0.0143853673690917\\
};
\addlegendentry{64QAM, NN}
\addplot [color=forestgreen, line width = 1.0pt, mark size= 2pt, mark=*, mark options={solid, fill = emerald, draw= forestgreen}]
 plot [error bars/.cd, y dir = both, y explicit,error bar style={line width=0.8pt,solid},
      error mark options={line width=0.8pt,mark size=3pt,rotate=90}]
 table[row sep=crcr, y error plus index=2, y error minus index=3]{%
1       4.97841661          0.00531843          0.00531843\\
1.25	5.00757671261714	0.00610502182275793	0.00610502182275793\\
1.5	5.03081818177387	0.00726516492511075	0.00726516492511075\\
1.75	5.04890922494249	0.00792150972462659	0.00792150972462659\\
2	5.06030257863442	0.00834563180628463	0.00834563180628463\\
2.25	5.06426431366264	0.00841853525831069	0.00841853525831069\\
2.5	5.06124491880202	0.0100652452807466	0.0100652452807466\\
2.75	5.04999035163232	0.0116270742443903	0.0116270742443903\\
3	5.03129532225953	0.0122980058074035	0.0122980058074035\\
};
\addlegendentry{memoryless GS, NN}

\addplot [color=mycolor1, line width = 1.0pt, mark size= 2.5pt, mark=diamond*, mark options={solid, fill = ballblue, draw = mycolor1}]
 plot [error bars/.cd, y dir = both, y explicit,error bar style={line width=0.8pt,solid},
      error mark options={line width=0.8pt,mark size=3pt,rotate=90}]
 table[row sep=crcr, y error plus index=2, y error minus index=3]{%
1.00    5.0147  0.00567537  0.00567537\\
1.25	5.0475	0.00572748	0.00572748\\
1.5	    5.0739	0.00560916	0.00560916\\
1.75	5.0956	0.00655706	0.00655706\\
2	    5.1108	0.00794921	0.00794921\\
2.25	5.12599121	0.00904225	0.00904225\\
2.5	    5.13193448	0.00942703	0.00942703\\
2.75	5.12969263	0.00893733	0.00893733\\
3	    5.12196140	0.01054046	0.01054046\\
};
\addlegendentry{GS + pre-emph., NN}

\end{axis}
\draw[arrows=<->,line width=.7pt](0.345\textwidth,0.18\textwidth)--(0.345\textwidth,0.29\textwidth);
\node[] at (0.3156\textwidth,0.24\textwidth) {\footnotesize{$\approx$ 0.11}};


\draw[arrows=<->,line width=.7pt](0.393\textwidth,0.18\textwidth)--(0.393\textwidth,0.36\textwidth);
\node[] at (0.425\textwidth,0.24\textwidth) {\footnotesize{$\approx$ 0.18}};

\node[] at (0.16\textwidth,0.35\textwidth) {\small{(a)}};

\end{tikzpicture}%
    \vspace{-0.3cm}
%
%
\pgfplotsset{every tick label/.append style={font=\footnotesize}}
\definecolor{darkblue}{rgb}{0.00000,0.44700,0.74100}%

\begin{tikzpicture}
\begin{axis}[%
width=0.25\textwidth,
height=0.25\textwidth,
at={(-0.5in,0.0in)},
xlabel={I},
ylabel={Q},
xlabel style={font=\footnotesize \color{white!15!black},at={(0.5,-2ex)}},
ylabel style={font=\footnotesize \color{white!15!black},at={(-2ex,0.5)}},
]
\addplot [color=darkblue, mark=*, only marks,mark size=1pt, mark options={solid,fill=darkblue, draw=darkblue}]
  table[row sep=crcr] {
-0.863	-0.6445	\\
-0.0328	-0.3049	\\
0.1551	0.1783	\\
0.8034	-1.2924	\\
0.3422	-1.3782	\\
0.8658	-0.8971	\\
0.8014	0.5509	\\
-0.9917	-0.2889	\\
0.2055	0.7292	\\
0.5104	0.3761	\\
-0.0818	0.1013	\\
0.9187	-0.0437	\\
-0.0401	0.3773	\\
0.5066	-0.9789	\\
-1.2431	0.4416	\\
0.3863	0.1003	\\
0.1691	-0.9977	\\
1.1218	-0.2812	\\
-0.4285	-0.0306	\\
1.0068	0.2763	\\
0.2526	-0.2855	\\
-0.1384	-0.1091	\\
0.5835	-0.6271	\\
0.1742	-0.0804	\\
0.0059	0.9915	\\
-0.3098	0.8423	\\
0.5128	0.72	\\
1.2082	0.6352	\\
-0.9513	0.8817	\\
0.7346	-0.2992	\\
-1.2613	-0.6267	\\
0.9043	-0.5576	\\
0.545	-0.0828	\\
1.3555	0.1438	\\
0.307	-0.6886	\\
0.4587	-0.3602	\\
-0.0434	-0.7878	\\
1.4943	-0.3033	\\
-0.3729	0.48	\\
0.2502	0.439	\\
-0.6141	0.6926	\\
-0.4212	-0.5621	\\
0.1284	-0.5258	\\
-0.0849	0.6231	\\
-0.4675	-1.288	\\
0.6986	0.1695	\\
-0.0017	1.3654	\\
-0.2821	-0.942	\\
0.4203	1.1014	\\
-0.0593	-1.2815	\\
1.2589	-0.7622	\\
-0.601	0.2489	\\
-0.8641	-1.0641	\\
0.8505	0.973	\\
-0.9869	0.1044	\\
-0.3551	-0.2795	\\
-0.1647	-0.5342	\\
-0.6544	-0.3635	\\
-1.3447	-0.0842	\\
-0.8501	0.4413	\\
-0.708	-0.0561	\\
-0.5673	-0.8151	\\
-0.501	1.1588	\\
-0.3015	0.2164	\\
    };
\end{axis}
\begin{axis}[%
width=0.25\textwidth,
height=0.25\textwidth,
at={(0.98in,0.0in)},
xlabel style={font=\footnotesize \color{white!15!black},at={(0.5,-2ex)}},
ylabel style={font=\footnotesize \color{white!15!black},at={(-2ex,0.5)}},
xlabel={I},
ylabel={Q},
]
\addplot [color=darkblue, mark=*, only marks,mark size=1pt, mark options={solid,fill=darkblue, draw=darkblue}]
  table[row sep=crcr] {
0.66695179	-0.55399087\\
-0.98021194	0.60515884\\
0.89556588	-0.29240531\\
-0.43580670	-0.47303627\\
-0.19086790	1.09731701\\
-0.18570459	0.79322621\\
1.58283851	-0.00370822\\
-1.18835466	-0.22732896\\
-0.80337227	-0.40345266\\
0.12278750	-0.32092248\\
0.04881853	-0.88513788\\
-0.63425522	-0.74071803\\
0.17598671	0.40467247\\
-0.87133321	0.22783986\\
0.47665811	-0.78435102\\
0.01969721	1.46481523\\
0.19082873	0.74351379\\
1.21429297	-0.22164576\\
-0.22740455	-0.17960319\\
-0.49834112	-0.16709736\\
-0.78627489	-0.06329299\\
-0.10132570	-1.30692229\\
0.48036999	0.02114218\\
0.76710129	0.03985373\\
0.71931479	0.64857263\\
0.73635797	-1.05390103\\
-1.27952273	0.24608568\\
0.42961556	0.53588196\\
0.49284327	0.89101851\\
1.18578098	-0.99779614\\
0.01448982	-0.06174777\\
0.92657079	0.35828325\\
0.83331013	1.04951904\\
0.18211960	1.06448922\\
-0.17394801	-0.65886870\\
-0.31801152	-0.95029031\\
-0.26390264	0.08501006\\
0.40084275	-0.41592777\\
-0.52862155	1.30061418\\
-0.51105574	0.17573833\\
1.40980494	-0.55234678\\
1.08076470	0.08515604\\
0.50743922	1.34403438\\
-0.33903561	0.51447191\\
0.61654641	0.31071022\\
-0.65683915	-1.18679963\\
-0.62091535	0.48625930\\
0.97116708	-0.65696079\\
-0.16059918	0.30980576\\
1.28609220	1.02107819\\
0.19000642	-0.60882841\\
0.45791418	-1.44360647\\
-0.02061833	0.56495743\\
0.60107751	-0.22574994\\
0.06648342	0.15843463\\
1.36840944	0.39457522\\
-0.10876941	-0.40722137\\
0.29634172	-1.08487949\\
0.28394726	-0.12563109\\
-1.05819759	-0.75191673\\
-0.94663195	1.01249234\\
-0.55442540	0.84709478\\
1.07375745	0.67163967\\
0.33459845	0.20683819\\
    };
\end{axis}



\node[] at (-0.12in,1.2in) {\footnotesize{(b) without~pre-emph.~(memoryless) }};
\node[] at (-0.1in,1.05in) {\footnotesize{ @ 2.25~dBm }};

\node[] at (1.5in,1.2in) {\footnotesize{(c) with~pre-emph. }};
\node[] at (1.4in,1.05in) {\footnotesize{ @ 2.5~dBm }};

\end{tikzpicture}%
    \caption{(a) MI obtained by 64-QAM, learned GS-64 constellation and by joint GS-64 and nonlinear pre-emphasis. Examples of the learnt Constellations (b) memoryless GS-64 and (c) GS-64 $+$ pre-emphasis. }    
    \label{fig:MI}
\end{figure}
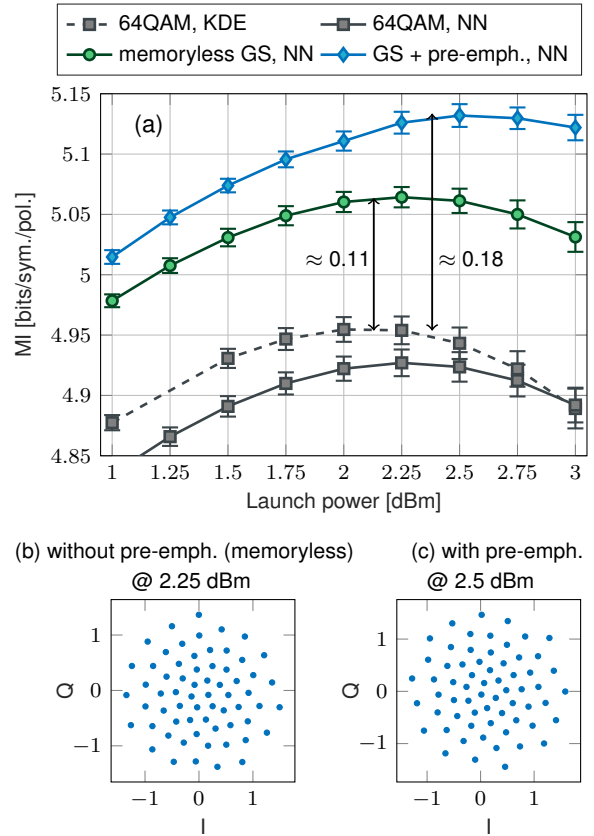

Fig.~\ref{fig:MI}(a) shows the E2E MI optimized for different input powers. We also plot the E2E MI optimized without pre-emphasis NN, neglecting the channel memory. For each point, the standard deviation over 10 simulation runs is also shown. 
As a reference, we plot the MI of standard 64-QAM without pre-emphasis, evaluated with two methods: by training the decoder NN to learn $P(x|y)$, and by using a 
kernel density estimator (KDE) to estimate $P(y|x)$. The latter gave larger values, highlighting the opportunities for decoder improvement, and is taken as a reference. 
We observe that the E2E learning results in a considerable gain. Without pre-emphasis, optimization of the constellation shaping gives MI gain of $\approx0.11$ bits/sym./pol. and with pre-emphasis, the MI gain increases further to $\approx0.18$ bits/sym./pol. while the optimal power is also increased by $\approx 0.5$~dB.

\section{Conclusion}
We presented a novel End-to-End learning approach optimizing geometric constellation shaping and a nonlinear pre-emphasis for coherent fiber-optic communications, resulting in a considerable mutual information increase in simulation. The proposed technique, relying on the ``parallelizable'' regular perturbation model, can be used for different fiber channels. 

\footnotesize
\textit{Acknowledgements}:
This project has received funding from EU Horizon 2020 program under the Marie Sk\l{}odowska-Curie grant agreement No. 766115 (FONTE). JEP is supported by Leverhulme Project RPG-2018-063. SKT acknowledges the support of EPSRC project TRANSNET.
\normalsize
\bibliographystyle{IEEEtran}
\bibliography{references_new.bib}

\end{document}